\documentclass[prb,aps,superscriptaddress,twocolumn,amssymb,showpacs]{revtex4-1}
\usepackage{graphicx}
\usepackage{longtable}
\usepackage{epsfig}
\usepackage{dcolumn}
\usepackage{bm}
\usepackage{color}
\definecolor{lightblue}{RGB}{173,216,230}
\begin{document}

\title{Strong Effects of Cation Vacancies
       on the Electronic and Dynamical Properties of FeO}

\author{     Urszula D. Wdowik  }
\affiliation{Institute of Technology, Pedagogical University,
             Podchor\c{a}\.zych 2, PL-30084 Krak\'{o}w, Poland }

\author{     Przemys\l{}aw Piekarz }
\affiliation{Institute of Nuclear Physics, Polish Academy of Sciences,
             Radzikowskiego 152, PL-31342 Krak\'ow, Poland }

\author{     Krzysztof Parlinski }
\affiliation{Institute of Nuclear Physics, Polish Academy of Sciences,
             Radzikowskiego 152, PL-31342 Krak\'ow, Poland }
	
\author{     Andrzej M. Ole\'{s} }
\affiliation{Marian Smoluchowski Institute of Physics, Jagellonian University,
             Reymonta 4, PL-30059 Krak\'ow, Poland}
\affiliation{Max-Planck-Institut f\"ur Festk\"orperforschung,
             Heisenbergstrasse 1, D-70569 Stuttgart, Germany}

\author{     J\'{o}zef Korecki }
\affiliation{Jerzy Haber Institute of Catalysis and Surface Chemistry,
             Polish Academy of Sciences, \\
             Niezapominajek 8, PL-30239 Krak\'ow, Poland }
\affiliation{Faculty of Physics and Applied Computer Science,
             AGH University of Science and Technology, \\
             aleja Mickiewicza 30, PL-30059 Krak\'ow, Poland}

\begin{abstract}

We report pronounced modifications of electronic and vibrational
properties induced in FeO by cation vacancies, obtained within density
functional theory incorporating strong local Coulomb interactions at
Fe atoms. The insulating gap of FeO is reduced by about 50\% due to
unoccupied electronic bands introduced by trivalent Fe ions stabilized
by cation vacancies. The changes in the electronic structure along
with atomic displacements induced by cation vacancies affect strongly
phonon dispersions via modified force constants, including those at
atoms beyond nearest neighbors of defects.
We demonstrate that theoretical phonon dispersions
and their densities of states reproduce the results of inelastic neutron
and nuclear resonant x-ray scattering experiments \emph{only} when Fe
vacancies and Coulomb interaction $U$ are both included explicitly in
\emph{ab initio} simulations, which also suggests that the
electron-phonon coupling in FeO is strong.
\end{abstract}
\date{\today}


\maketitle

As explained by the theory of Mott \cite{Mott} and Hubbard,
\cite{Hubbard} the insulating state in transition metal oxides results
from the electron localization in the narrow $3d$ band due to the local
Coulomb interactions.\cite{Ima98} Iron oxides are in the focus of
present-day research covering a broad range of complex phenomena such
as the Verwey transition in magnetite \cite{review} or structural
instabilities in polar nanosystems.\cite{Goniakowski}
High-resolution neutron and x-ray scattering experiments
on magnetite \cite{Wri01} suggest a considerably more complex charge
order than originally invoked by Verwey.\cite{Ver39} It has been shown
that the microscopic understanding of this phenomenon requires full
information about both the electronic structure and the lattice dynamics
in the presence of strong electron interactions which significantly
amplify the electron-phonon coupling, as predicted in the theory
\cite{Pie06} and confirmed recently by inelastic x-ray scattering.
\cite{Hoe12} This could be a common feature of iron oxides ---
indeed, the exchange-induced phonon splitting observed in w\"ustite
\cite{Kan12} suggests a strong coupling between electronic and
lattice degrees of freedom.

W\"ustite, Fe$_{1-x}$O, is another strongly correlated insulator within
iron oxides, with the insulating energy gap of $\sim 2.4$ eV arising
mainly from the O($2p$) $\rightarrow$ Fe($3d$) charge-transfer process.
\cite{Zaanen} Below the N\'{e}el temperature $T_N=198$ K
antiferromagnetic (AF) order sets in, which suggests that also here
strong on-site Coulomb interaction $U$ plays a role. A stable crystal
of Fe$_{1-x}$O shows large Fe-deficiency on the level $0.05<x<0.15$,
\cite{notex} which substantially influences its structural properties.
\cite{Roth1,Cheetham} It may also affect the geophysical processes in
the Earth's lower mantle where w\"ustite, existing in the (Mg,Fe)O
solid solution, is one of the most important constituents.
\cite{Ringwood} So far, first-principles investigations on a
vacancy-defected FeO \cite{Press} and
other transition metal oxides \cite{Kod03,Wdowik3} are scarce.
Previous studies were devoted to the stoichiometric FeO,
\cite{Anisimov90,Tran} leaving the effect of cation vacancies on the
electronic and lattice properties of w\"{u}stite largely unexplored.

At room temperature, Fe$_{1-x}$O crystallizes in a rock-salt structure
and exhibits the rhombohedral distortion with a weak elongation along
the [111] direction in the AF phase below $T_N$. The magnetic moments
on the Fe cations are aligned perpendicular to the ferromagnetically
ordered (111) planes reversing their orientations from one to another
plane.\cite{Schull} Due to the cation deficiency and crystal-field
effects, the average magnetic moment $m=3.32 \mu_B$ measured at 4.2~K
is significantly reduced from the ionic Fe$^{2+}$ value $m=4 \mu_B$
(for spin $S=2$).\cite{Roth2} Large splitting of phonon modes below
$T_N$ indicates strong spin-phonon coupling in w\"{u}stite.\cite{Kan12}

In this Rapid Communication we demonstrate that Fe vacancies
(V$_{\rm Fe}$) strongly modify the structural, electronic, and
vibrational properties of w\"{u}stite.
Our theoretical investigations uncover qualitative difference between
stoichiometric FeO and defected Fe$_{1-x}$O containing either 3\% or
6\% of cation vacancies. The present {\it ab initio} results for
Fe$_{1-x}$O are in good agreement with the available experimental data
from inelastic neutron scattering (INS) \cite{Kugel} and nuclear
resonant inelastic x-ray scattering (NRIXS),\cite{NIS} which validates
our findings.

Notably, a realistic description of the electronic structure of Mott
insulating states in transition metal monoxides requires methods
which implement explicitly local Coulomb interactions.
\cite{Sol08,Ima10} Previous theoretical studies within the local
density approximation (LDA) using the Coulomb interaction $U$ (LDA+$U$)
\cite{Anisimov,Lie95} confirmed the prominent role of local electron
interactions on the band structure of defect-free FeO and significant
contribution of the O($2p$) states at energies just below the Fermi
energy. Cation vacancies are responsible for charge
redistribution and appearance of trivalent Fe ions in the octahedral
and tetrahedral (interstitial) positions.\cite{Cheetham,Press}
The highly correlated nature of $3d$ electronic states has also a
strong impact on the lattice dynamics of transition metal oxides. The
experimental data have been well reproduced within the LDA+$U$ approach
\cite{Lie95} applied to NiO,\cite{Savrasov} CoO,\cite{Wdowik1} and
MnO;\cite{Wdowik2} further improvements were obtained in the Green's
function approach.\cite{Massidda}
Until now, lattice dynamics of stoichiometric FeO has only been studied
in a semiempirical model.\cite{Upa01}

The present calculations were performed within the generalized gradient
approximation (GGA) and the projector augmented-wave method \cite{PAW}
implemented in the {\sc vasp} code.\cite{vasp} The on-site interactions
at Fe ions are: the Coulomb element $U=6$~eV and Hund's exchange
$J=1$~eV.\cite{Tran,Anisimov} The AF supercell composed of 64 atoms was
used to calculate stoichiometric FeO. The Fe$_{1-x}$O crystals with
$x=3$\% and $x=6$\% were modeled by removing either one or two Fe atoms
from the optimized AF supercell. Various V$_{\rm Fe}$-V$_{\rm Fe}$
distances within the simulated supercell
with $x=6$\% were considered. Finally, a configuration with the lowest
energy, corresponding to the distance between two vacancies of 1.22$a$,
with $a$ being the lattice constant of the perfect FeO,
was chosen for further band structure and phonon calculations.
The energy difference between the lowest and the
next stable configuration amounts to 7 meV/atom indicating
rather high mobility (diffusion) of vacancies in w\"{u}stite.
The volumes of the supercells representing Fe$_{1-x}$O systems were
fixed at the optimized value of the vacancy-free lattice, the symmetry
constraints were removed, and all atoms were allowed to relax.

We identified two distinct valence states of Fe cations for each
Fe$_{1-x}$O composition. Majority of cations are Fe$^{2+}$ ions with
moments $m=3.72 \mu_B$, but there are two iron ions per each
V$_{\rm Fe}$ that have enhanced charges and moments
($m=4.22 \mu_B$). For single-vacancy system, they are located at
distances $a$ and $1.73a$ from V$_{\rm Fe}$.
The Bader charge analysis \cite{Bader} indicates an increase in their
valence charges by $+0.31e$ over the valence of Fe$^{2+}$ ions.
We assign the above features to the Fe$^{3+}$ ions.

\begin{figure}[t!]
\includegraphics[width=8.1cm]{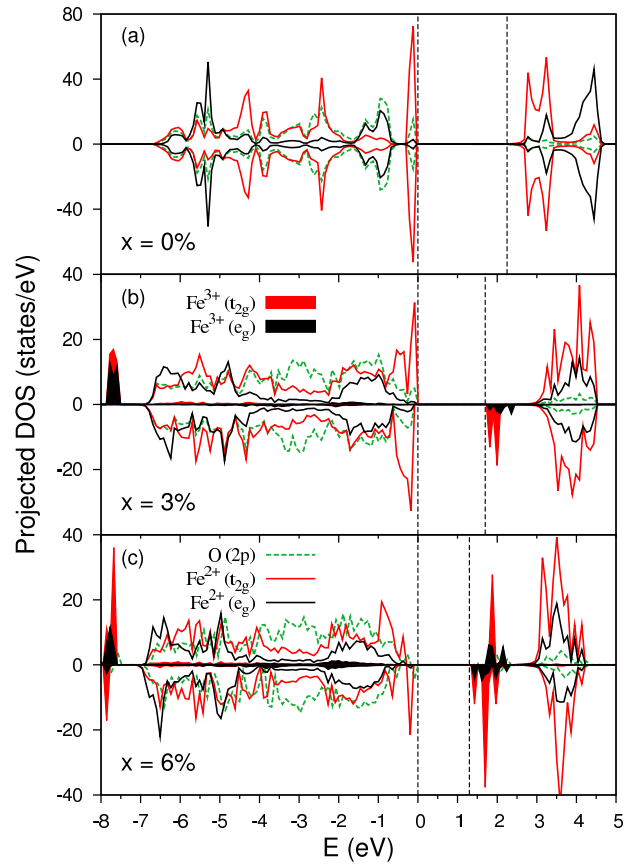}
\caption{(Color online)
Total orbital projected electronic DOS obtained for:
(a) the stoichiometric FeO, and defected Fe$_{1-x}$O, with
(b) $x=3$\%, and
(c) $x=6$\%.
The positive (negative) DOS represents the spin-up (spin-down) states
in each case; the gaps between the valence and conduction band are
indicated by dashed lines. Parameters: $U=6$~eV, $J=1$~eV.
}
\label{fig:fig1}
\end{figure}

The orbital projected electron densities of states (DOS) for FeO and
the two Fe$_{1-x}$O systems considered here are dominated by the O($2p$)
and Fe($3d$) states; the latter showing the crystal-field splitting
between the $t_{2g}$ and $e_g$ symmetry.
As in the previous studies,\cite{Anisimov90} the electronic
states close to the energy gap are occupied by the minority spin
$t_{2g}$ electrons, see Fig. \ref{fig:fig1}.
The empty Fe($3d$) states are split and shifted to higher energies due
to the Hubbard interaction $U$, while the O($2p$) states are almost
fully occupied and hybridize with the Fe($3d$) states in a broad energy
range below the Fermi energy.

We observe significant modifications of the electronic
structure for the system with incorporated cation vacancies.
First of all, the projected electronic DOS becomes asymmetric in the
distribution of the spin-up and spin-down components of the $3d$
states after removing one Fe cation from the spin-down sublattice.
Also, unlike in the stoichiometric FeO, see Fig.~\ref{fig:fig1}(a),
the oxygen bands of the w\"{u}stite crystals with different
concentration of Fe vacancies remain spin-polarized. Each V$_{\rm Fe}$
induces changes in the charge distribution of its immediate
neighborhood, pushing the O($2p$) states closer to the Fermi energy.
As the most interesting feature we identify additional electronic
bands that originate from the $3d$ states of Fe$^{3+}$ ions.
In Fe$_{1-x}$O with $x=3$\%, the occupied $3d$ states of Fe$^{3+}$
ions form a narrow band at about $-7.6$~eV, see Fig.~\ref{fig:fig1}(b).
It lies below the main manifold of the Fe$^{2+}$ states and consists of
all five Fe($3d$) states. The unoccupied Fe$^{3+}$($3d$) states appear
as the lowest energy empty states and reduce the insulating energy gap
of w\"{u}stite from $\sim 2.2$~eV to $\sim 1.8$~eV.\cite{noteCoO}
Spin polarization of the Fe$^{3+}$ states depends on the Fe (spin-up or
spin-down) sublattice into which the
V$_{\rm Fe}$ is introduced. The electronic bands due to the trivalent
Fe ions in the Fe$_{1-x}$O system with $x=6$\% exist in both spin
channels as the cation vacancies occupy both spin-sublattices. A larger
concentration of vacancies $x=6$\% broadens electronic bands and leads
to further decrease of the
insulating gap to the value of $\sim 1.3$~eV \cite{noteuj}.
This value is quite close to the observed optical gap of $\sim 1.0$
eV.\cite{Kan12} The electronic states induced by disorder are
expected to influence transport properties of FeO.\cite{notepo}

\begin{figure}[t!]
\includegraphics[width=8.0cm]{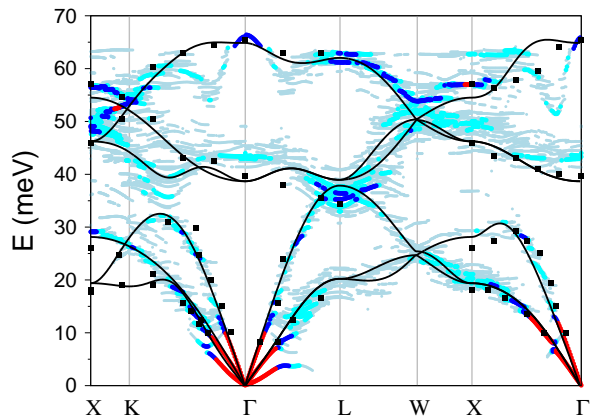}
\caption{(Color online) Phonon dispersion relations for stoichiometric
FeO (solid lines) and the relative intensities of the phonon modes in
Fe$_{1-x}$O with $x=3$\%, for $U=6$~eV, $J=1$~eV. The experimental data
(solid squares) are adopted from Ref. \cite{Kugel}.
The highest intensity mode is taken as a reference (100\%);
relative intensities are arranged into the following ranges indicated by 
the intensity of gray scale (color):
5-20\% (\textcolor{lightblue}),
20-50\% (\textcolor{cyan}),
50-80\% (\textcolor{blue}),
80-100\% (\textcolor{red}).
High-symmetry points are
(in units of $2\pi/a$): $\Gamma=(0,0,0)$,
$X=(\frac12,\frac12,0)$, $K=(\frac38,\frac38,\frac34)$,
$L=(\frac12,\frac12,\frac12)$, $W=(\frac12,\frac14,\frac34)$.
}
\label{fig:fig2}
\end{figure}

Phonons in FeO and Fe$_{1-x}$O were determined by the
direct method,\cite{direct} where the Hellmann-Feynman forces were
obtained by displacing all symmetry non-equivalent atoms from their
equilibrium positions with the amplitude of 0.03~\AA{} and applying
both positive and negative displacements to minimize systematic error.
\cite{vasp} The transverse optic (TO) phonon modes followed
directly from the diagonalization of the dynamical
matrix,\cite{direct} while the longitudinal optic (LO) modes were
obtained by introducing into the dynamical matrix the non-analytical
term,\cite{Pick} the latter dependent on the Born effective charge
tensor $\mathbf{Z}^{*}$ and the electronic part of dielectric constant
$\mathbf \epsilon_{\infty}$. Both $\mathbf{Z}^{*}$ and
$\mathbf \epsilon_{\infty}$ were determined within the linear response
method \cite{Gajdos} implemented in the {\sc vasp} code.
Our calculations resulted in $|Z_{\rm Fe}^*|=|Z_{\rm O}^*|=2.2$ and
$\epsilon_{\infty}=5.262$.\cite{notez*}
$\epsilon_{\infty}$ and $|Z^*|$ averaged over particular
sublattices remain almost insensitive to the incorporated cation
vacancies,\cite{BornEffChgr} whereas the Fe$^{3+}$ ions acquire the
effective dynamical charge of $+1.3$ over that for Fe$^{2+}$ ions.

Imposed cubic symmetry constraints result in 6 phonon branches. The
transverse acoustic (TA) and TO phonon modes remain doubly degenerate
along the $\Gamma$-$X$ and $\Gamma$-$L$ directions. Despite an apparent
correlation between our theoretical curves and the INS data,\cite{noteu}
we notice some discrepancies for the TO branch close to the $L$ point
and the longitudinal acoustic (LA) branch near the $X$
point, see Fig.~\ref{fig:fig2}. An evident departure of the calculated
LO branch from the experimental points along the $\Gamma - X$ direction
is mainly due to the interpolation function used here to calculate the
LO-TO splitting.\cite{direct}

\begin{figure}[t!]
\includegraphics[width=8.0cm]{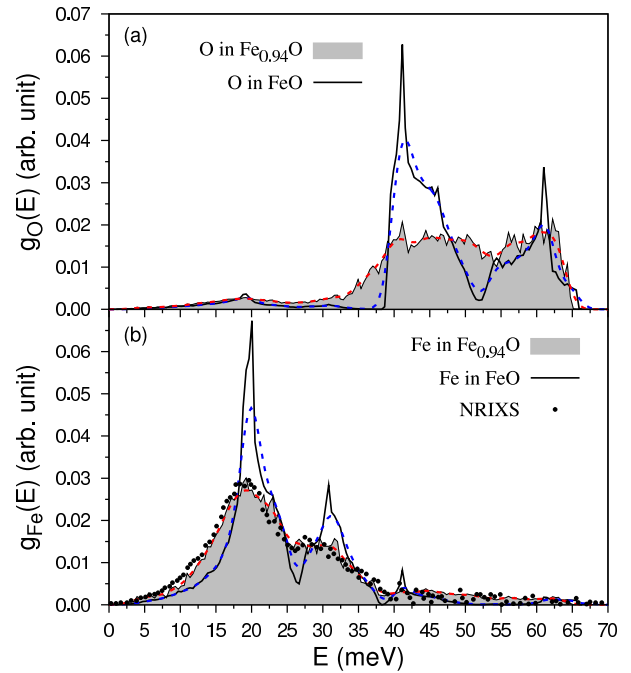}
\caption{(Color online) Partial phonon DOS obtained for stoichiometric
FeO (solid lines) and Fe$_{1-x}$O with $x=6$\% (shaded area) for
sublattice of:
(a) oxygen and
(b) iron ions.
Experimental points in (b) are taken from the NRIXS experiment.\cite{NIS}
Dotted lines denote the Gaussian convolutions of our theoretical
results with the experimental resolution of 2.4~meV.
Parameters: $U=6$~eV, $J=1$~eV.}
\label{fig:fig3}
\end{figure}

Some of the above discrepancies can be explained by considering phonons
in the FeO crystals defected by cation vacancies. However, we have to
remark that an elaboration of the resulting phonon dispersion curves in
the system with point defects requires a special treatment which
involves application of the phonon form factor.\cite{Sjolander}
We follow the same methodology as used before for the strongly
correlated CoO with various point defects
(cation vacancies and Fe impurities).\cite{Wdowik3}
The resulting intensities of the phonon modes in Fe$_{1-x}$O
with $x=3$\% are overlaid onto the phonon dispersions for the
stoichiometric FeO in Fig.~\ref{fig:fig2}.
Discontinuities in some phonon branches come from the 5\% cut-off
applied to the relative intensities of the phonon modes in Fe$_{1-x}$O.
Although the calculated phonon spectrum is dominated by the modes of
low relative intensities, the INS data concentrate close to the modes
with intensities exceeding 30\%. Moreover, the calculated phonons at
the Brillouin zone boundaries, e.g. at the $L$ and $X$ points,
match better the experimental data.
Finally, despite many similarities between the phonon spectra of FeO
with 3\% and 6\% vacancies, the system with doubled vacancy concentration
exhibits significantly more phonon modes carrying low intensities.

\begin{figure}[t!]
\includegraphics[width=8.0cm]{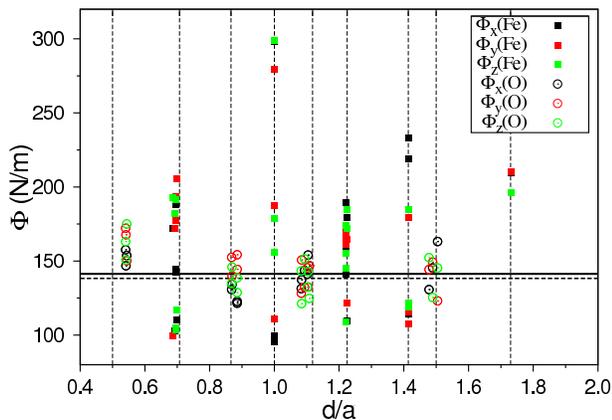}
\caption{(Color online)
Diagonal elements $\Phi$ of the on-site force-constants matrix for
distance $d/a$ from the vacancy in the Fe$_{1-x}$O system with $x=3$\%.
Vertical dashed lines denote ideal positions of the Fe and O atoms,
whereas the horizontal solid and dashed lines stand for the
respective force constants in the stoichiometric FeO.
Parameters: $U=6$~eV, $J=1$~eV.}
\label{fig:fig4}
\end{figure}

Essential differences in the dynamics of the stoichiometric and
defected lattices are revealed by the phonon DOS, see Fig.
\ref{fig:fig3}. Here, our theoretical results of partial Fe DOS obtained
for stoichiometric FeO and Fe$_{1-x}$O with $x=6$\% are compared to the
experimental data of NRIXS \cite{NIS} performed on the w\"{u}stite
sample with the nonstoichiometry of 5.3\%, see Fig.~\ref{fig:fig3}(b).
The iron and oxygen contributions span different energy ranges which
overlap weakly due to large difference between their masses.
Both convoluted and bare theoretical phonon spectra of stoichiometric
FeO show quite narrow and well-resolved peaks, opposite to the broad
phonon maxima in the Fe$_{1-x}$O spectra. These latter spectra agree
remarkably well with the experimental results.\cite{NIS} A smearing of
the two-peak structure, inevitably connected with the presence of Fe
vacancies, becomes even more enhanced for the O-sublattice, indicating
its greater sensitivity to the changes in the local crystal environment.

To clarify the origin of changes in atomic vibrations,
we have analyzed the force constant matrix for the w\"{u}stite crystal
with $x=3\%$, see Fig. \ref{fig:fig4}. The on-site force constants
$\Phi_{\alpha}$ along the main crystallographic axes $\alpha=x,y,z$
increase for the vacancy neighboring oxygen atoms over the
respective ones in the idealized vacancy-free FeO.
They are displaced outward the vacancy by $\sim0.18$~\AA. On the other
hand, the O atoms occupying the subsequent shells are attracted toward
the nearest Fe atoms and their force constants remain equally
distributed around the value corresponding to the perfect lattice.
Similar but weaker effects on force constants
have also been encountered in the cation-deficient CoO.\cite{Wdowik3}

Relaxations of the atomic positions in the Fe-sublattice are less
pronounced, except the second nearest neighbors to a vacancy
which shift closer to it.\cite{Schweika} The force
constants at the iron sites show considerably broader distribution
range than those at oxygens sites. Moreover, the force
constants at Fe$^{3+}$ ions are significantly enhanced from their
values at Fe$^{2+}$ ions, with the largest values of $\Phi_{\alpha}$
at the distance $a$ from V$_{\rm Fe}$.
This clearly demonstrates a strong impact of the w\"ustite
electronic structure on its interatomic forces and accounts for the
observed broadening of peaks in the phonon DOS. It may also explain
large spectral linewidths of the dielectric loss function and strong
electron-phonon coupling observed
recently for FeO with the nonstoichiometry of 7-8\%.\cite{Kan12}
Such pronounced changes in the phonon DOS, in particular the excess of
spectral intensity at lowest energies, influence also thermodynamic
properties of FeO. For instance, in case of one cation vacancy the
calculations predict the increase of lattice heat capacity at $T=50$ K
by about $10 \%$ over the stoichiometric FeO, and enhanced
thermal atomic displacements in the neighborhood of a vacancy.

Summarizing, we have reported dramatic changes of the electronic
structure and vibrational spectra of w\"{u}stite caused by cation
vacancies. Significant charge redistribution on Fe ions induces the
empty electronic states above the valence band and strongly reduces
the insulating gap. The changes in the electronic structure and atomic
displacements generated by defects substantially modify the force
constants not only in the immediate neighborhood of vacancies but
also at more distant atoms. It results in broadening of peaks in the
phonon spectra, in close agreement with available experimental data.
Predicted changes in the dynamics of the oxygen sublattice in
Fe$_{1-x}$O provide an experimental challenge that could be
resolved, for example, by incoherent INS experiments.

{\it Acknowledgments.---}
P. Piekarz and A.M. Ole\'s kindly acknowledge financial support
by the Polish National Science Center (NCN) under Projects
No. 2011/01/M/ST3/00738 and No. 2012/04/A/ST3/00331.


\begin{thebibliography}{99}
\bibitem{noteCoO} The electronic states are localized inside the insulating
                gap also in the vacancy-defected CoO,\cite{Wdowik1}
                in stark contrast to MnO and NiO where cation vacancies
                induce only shallow states.\cite{Kod03}
\bibitem{noteuj} Estimates imply that the Mott-Hubbard gap closes
                 at $U<4$~eV for each composition.
\bibitem{notepo} The coupling between electrons and lattice distortions
                in defected structure may be responsible for polaronic
                character of charge carriers observed at low
                temperatures.\cite{Kan12}
\bibitem{notez*} The small rhombohedral distortions of the stoichiometric
                 FeO lattice were neglected and under such an assumption
                 both $\mathbf{Z}^{*}$ and $\mathbf\epsilon_{\infty}$
                 tensors could be described by single parameters $|Z^*|$
                 and $\epsilon_{\infty}$, respectively.
\bibitem{BornEffChgr} $\epsilon_{\infty}=5.261$ and
                 $\epsilon_{\infty}=5.483$ were obtained for the
                 Fe$_{1-x}$O systems with $x=3$\% and $x=$6\%, respectively.
                 Averaged $|Z_{\rm Fe}^*|=2.3$ and $|Z_{\rm Fe}^*|=2.4$ are
                 found for Fe ions in the systems with $x=3$\% and $x=6$\%,
                 respectively. No changes were observed for average
                 O charge $|Z_{\rm O}^*|$.
\bibitem{noteu} Such a good agreement between the experimental data and
                the calculated phonon dispersions could be obtained only
                with $U\simeq 6$ eV and $J\simeq 1$ eV; at $U=J=0$ one
                finds much poorer agreement including soft mode behavior.
\end{thebibliography}
\end{document}